\documentclass{elsart}
\usepackage{adnd}
\usepackage{longtable}
\usepackage{lscape}

\usepackage{epstopdf}

\usepackage{hyperref}
\usepackage{mathptmx}

\usepackage{amsmath}

\usepackage{amssymb}

\usepackage[square,sort&compress]{natbib}
\bibpunct[]{[}{]}{,}{n}{}{;}
\usepackage{graphicx}

\begin{document}
\begin{frontmatter}
\journal{Atomic Data and Nuclear Data Tables}
\title{Comment on ``Atomic mass compilation 2012" by B.~Pfeiffer, K.~Venkataramaniah, U.~Czok, C.~Scheidenberger}

   \begin{center}
   {\normalsize
                G.~Audi$^{\mathrm{a}}$\hspace{-4pt}
                         \footnote{Corresponding author.\ \ \
                         {\it E-mail address:} amdc.audi@gmail.com (G. Audi).},
               K.~Blaum$^{\mathrm{b}}$,
               M.~Block$^{\mathrm{c}}$,
              G.~Bollen$^{\mathrm{d}}$,
            F.~Herfurth$^{\mathrm{c}}$,
             S.~Goriely$^{\mathrm{e}}$,
             J.C.~Hardy$^{\mathrm{f}}$,
            F.G.~Kondev$^{\mathrm{g}}$,
            H.-J.~Kluge$^{\mathrm{c,h}}$,
              D.~Lunney$^{\mathrm{a}}$,
           J.M.~Pearson$^{\mathrm{i}}$,
              G.~Savard$^{\mathrm{g}}$,
              K.~Sharma$^{\mathrm{j}}$,
                M.~Wang$^{\mathrm{k}}$,
             Y.H.~Zhang$^{\mathrm{k}}$
                  }\\ \vspace {2mm}
   {\small
      $^{\mathrm{a}}$
                  CSNSM, CNRS/IN2P3, Universit\'e Paris-Sud, B\^at. 108, F-91405 Orsay Campus, France
                  \\
      $^{\mathrm{b}}$
                  Max-Planck-Institut f\"ur Kernphysik,
                  Saupfercheckweg 1,
                  D-69117 Heidelberg, Germany
                  \\
      $^{\mathrm{c}}$
                  GSI Helmholtzzentrum f\"ur Schwerionenforschung GmbH,
                  Planckstrasse 1, D-64291 Darmstadt, Germany
                  \\
      $^{\mathrm{d}}$
                  National Superconducting Cyclotron Laboratory,
                  Michigan State University, East Lansing, MI 48824, USA
                  \\
      $^{\mathrm{e}}$
                  Institut d'Astronomie et d'Astrophysique, CP-226,
                  Universit\'e Libre de Bruxelles, 1050 Brussels, Belgium
                  \\
      $^{\mathrm{f}}$
                  Cyclotron Institute,
                  Texas A\&M University, College Station, TX 77843, USA
                  \\
      $^{\mathrm{g}}$
                  Argonne National Laboratory,
                  9700 S. Cass Avenue, Argonne, IL 60439, USA
                  \\
      $^{\mathrm{h}}$
                  University of Heidelberg, D-69120 Heidelberg, Germany
                  \\
      $^{\mathrm{i}}$
                  D\'epartement de Physique, Universit\'e de Montr\'eal,
                  Montr\'eal, Qu\'ebec, H3C 3J7, Canada
                  \\
      $^{\mathrm{j}}$
                  Department of Physics and Astronomy,
                  University of Manitoba, Winnipeg, MB R3T 2N2, Canada
                  \\
      $^{\mathrm{k}}$
                  Institute of Modern Physics, CAS,
                  509 Nanchang Rd., Lanzhou 730000, China
   }
\end{center}

\date{submtd 13.12.2013; rev. 05.05.2014}

\end{frontmatter}

   For more than half a century the Atomic Mass Evaluation ({\sc Ame}) has
striven to provide a consistent and comprehensive set of atomic masses (see
Ref. \cite{Ame2003a,Ame2003b,Ame2012a,Ame2012b} and references therein).
   Masses are measured directly by mass spectrometry techniques or deduced
from energy measurements in nuclear decays and reactions.
   In all cases, mass relations are established between two or more
nuclides, thus resulting in a large number of links that are meticulously
evaluated to ultimately obtain the masses for all known nuclei using a
least-squares-fit approach to all available experimental data.
   Those carefully recommended values play a seminal role in fundamental
research in many areas of natural sciences (chemistry, physics,
astrophysics, material physics, etc) and in an increasingly large number of
applications.
   For example, the 2003 edition {\sc Ame2003} \cite{Ame2003a,Ame2003b} has
been cited more than 2150 times according to Web of Science.
   The most recent evaluation, {\sc Ame2012} \cite{Ame2012a,Ame2012b}, and
the associated evaluation of nuclear properties, {\sc Nubase2012}
\cite{Nub2012}, were published in the December 2012 issue of the journal
Chinese Physics~C by a collaboration of scientists from Europe, China and
the USA.

   The {\sc Ame} approach and its long history are in contrast with the recent
publication of the so-called Atomic Mass Compilation (AMC12) \cite{Amc12},
which came to our knowledge only with its online publication on September 6,
2013.
   We, as regular users of, or contributors to, the {\sc Ame} mass tables, wish
to highlight several major differences between AMC12 and the traditional
{\sc Ame} series, by presenting below a few illustrative examples.
   It is not the purpose of this ``Comment" to present an extensive list of
all discrepancies between the AMC12 and the {\sc Ame2012} mass values nor to provide
detailed comments about all observed deviations.
   The reader might want to explore the original {\sc Ame2012} publication
\cite{Ame2012a}
where detailed comments are given on how each of the recommended atomic
mass values was derived.
   The consistency of the {\sc Ame2012} approach is also largely demonstrated
there, which is definitely not the case for the AMC12 compilation \cite{Amc12}.

   As a general rule, any credible evaluation must contain an extensive
compilation of all available experimental data.
   Indeed such a compilation is a prerequisite of the {\sc Ame} process and
it has already been published in the latest {\sc Ame2012}
\cite{Ame2012a,Ame2012b}.
   In this respect AMC12 does not add anything new to what has already been
published.
   Furthermore, although it is certainly necessary, a compilation is not
sufficient to provide a consistent set of values and this is exactly where
the AMC12 falls well short of {\sc Ame2012}.
   In the {\sc Ame} process, each piece of data is expressed as a linear
relation where the masses are treated as unknown parameters.
   The ensemble of relations is solved by the least-squares method, which
requires inversion of the associated normal matrix.
   One obvious advantage of the {\sc Ame} matrix approach over the simple
averages used in AMC12 is that it employs an overdetermined data set, which
allows for the evaluation of consistencies (or conflicts).

   By contrast, the method used in AMC12 is to combine any new result with
the mass values from the previous {\sc Ame2003}, independent of any other
new result, even if it is reported in the same paper.
   Given that many data have changed - and been improved - since 2003, this
procedure can only yield results that are less precise and, even worse,
less accurate.
   Correlations are ignored in AMC12, even though the {\sc Ame} has proven
that they are essential, given the strong entanglement of much of the input
data.
   Furthermore, in AMC12 the individual pieces of experimental data (decay
and/or reaction energies, etc.) are not given, but instead they have been
converted into mass-excesses, preventing easy access to the originally
published values.

   As a final remark, we urge the readers and users of mass tables not to
be misled by the similarity of the titles of ``The {\sc Ame2012} atomic
mass evaluation" \cite{Ame2012a,Ame2012b} with the AMC12 presented as
``Atomic mass compilation 2012" \cite{Amc12}.
   We would like to stress that the AMC12 is by no means the continuation
or an update of the work initiated by A.H. Wapstra in the 1950's.
   The methods used in the {\sc Ame} series have consistently proved
themselves and have led to useful and reliable tables of atomic masses from
the ensemble of experimental data obtained since 1934, and from all
laboratories around the world.
   The {\sc Ame2012} and the past {\sc Ame} evaluations were all endorsed
by the C2 Commission on Symbols, Units, Nomenclature, Atomic Masses and
Fundamental Constants of the International Union of Pure and Applied
Physics ({\sc Iupap}) whose aim is to promote international agreements on
the use of symbols, units, nomenclature and standards.

   The {\sc Ame} is a coordinated effort that involves collaboration
among several scientists from around the world.
   They make no claim to a monopoly on the world's mass data.
   However, because of the comprehensive approach used by the {\sc Ame}
collaboration, we consider the {\sc Ame} of superior accuracy to the more
simplistic AMC12 compilation.
   We hope that any confusion between the {\sc Ame} and the so-called AMC12
will not interfere with the increasing use of mass data for improving mass
models and in critical applications in the area of nuclear energy, and
elsewhere.

\section*{illustrative examples}

   \begin{enumerate}
   \item
   the unit for eV used in AMC12 is not the eV$_{90}$ for which full
explanation is given in the {\sc Ame}.
   This results in slight differences for many very precise masses.
   \item
   $^{45}$Cr : AMC12 missed the isomer at 100\,keV.
   \item
   $^{47}$Ar : AMC12's mass is --25910(100)\,keV compared to
--25210(90)\,keV ({\sc Ame2012}).
   Apparently the authors of the AMC12 missed the noteworthy result
obtained in 2006 by Gaudefroy et al., \cite{06Ga28}.
   \item
   $^{65}$As : AMC12 averages a theoretically estimated (via calculated
Coulomb Displacement Energy) value with an experimental result.
   Similarly $^{66}$Se and $^{69}$Br are given as being measured, which is
not true.
   \item
   $^{73}$Ge : AMC12 gives a precision of 1.6\,keV, whereas it is known now
with a precision of 0.06\,keV, i.e. 27 times better.
   The reason is that the AMC12 method is not able to combine the new very
precise mass of $^{74}$Ge with the $Q$-value for the (n,$\gamma$) reaction
on $^{73}$Ge.
   \item
   $^{100}$Sn : its value has changed significantly due to a very important
experiment at GSI \cite{12Hi07} published in Nature in June 2012, but
submitted already in October 2011, and accepted in April 2012.
   Since one author's name is common to both the Nature article and AMC12,
the authors of the latter must have been aware of this result in time to
include it in AMC12, something they should have done in light of its
importance and the significant impact it has on the mass of $^{100}$Sn.
   Obviously, the authors of AMC12 could not include this important result
since it is not directly a mass but requires the combination of Q-values!
   As a result AMC12's value for the mass of $^{100}$Sn is
--56780(710)\,keV instead of --57280(300)\,keV.
   \item
   $^{286}$Ed with $Z=113$ (called $^{286}$Uut in AMC12) : mass excess is
168202(896)\,keV in AMC12 and 169730\#(670\#) in {\sc Ame2012}. The difference
is thus 1520\,keV. Even worse, the $Q_{\alpha}$ value deduced from the
AMC12 table is 10012\,keV, whereas it is known from experiment to be
9770(100)\,keV.
   Strangely enough, since the mass of $^{270}$Db is not listed in the
AMC12, one can naively wonder how the mass of $^{286}$Ed, its precursor in
the $\alpha$-decay chain, could be determined.

   \end{enumerate}

\end{document}